\documentclass{article}

\makeatletter
\renewcommand{\@oddfoot}{--- ~\today~ ---\hfil
\raisebox{0.3ex}{\tiny --- ~D.~A.~Arbatsky~ ``What is
``Relativistic Canonical Quantization''?''~
---}\hfil --- ~\thepage~ ---}
 \makeatother

\begin{document}


\title{What is ``Relativistic Canonical Quantization''?}
\author{D.~A.~Arbatsky\footnote{ http://daarb.narod.ru/ , http://wave.front.ru/ }}
\date{\today}
\maketitle

\begin{abstract}
The purpose of this review is to give the most popular description
of the scheme of quantization of relativistic fields that was
named {\it relativistic canonical quantization} (RCQ). I do not
give here the full exact account of this scheme. But with the help
of this review any physicist, even not a specialist in the
relativistic quantum theory, will be able to get a general view of
the content of RCQ, of its connection with other known approaches,
of its novelty and of its fruitfulness.
\end{abstract}

\section*{Invariant Hamiltonian formalism}

It is generally known that for construction of quantized fields it
appears to be useful to calculate Poisson brackets of field
values. And a Poisson bracket is the notion of Hamiltonian
formalism.

If we have a problem to create a completely relativistic-invariant
scheme of quantization of fields, then we wish, first of all, to
formulate Hamiltonian formalism in relativistic-invariant form.

Science solved this problem for surprisingly long time. In fact,
it was generally accepted that Hamiltonian formalism can not be
formulated in explicitly relativistic-invariant form.

Nevertheless, with development of methods of symplectic geometry,
it became possible to formulate Hamiltonian formalism on the basis
of such notions that turned out to have relativistic-invariant
analogs. Here these notions are: phase space, symplectic structure
on phase space, canonical action of the one-parameter group of
time shifts in phase space.

For relativistic fields the analogs of these notions are,
correspondingly: invariant phase space, symplectic structure on
invariant phase space, canonical action of the Poincare group on
invariant phase space.

\paragraph{Invariant phase space.}
A point of usual phase space describes the dynamical state of
system in the fixed moment of time. If for each initial state of
the system the equations of motion are solvable, and uniquely,
then we can speak about one-to-one correspondence between the
phase space of the system (for the fixed moment of time) and the
set of solutions of the equations of motion. This set of solutions
of the equations of motion is called {\it invariant phase space}.

Invariant phase space possesses natural structure of manifold. If
dynamical system is a relativistic field, then as possible
coordinate functions on invariant phase space we can use values of
magnitude of the field in fixed points of space-time.

So, from the viewpoint of structure of invariant phase space,
possible values of field in a fixed point of space-time are just
one of the huge set of functions on invariant phase space.

\paragraph{Symplectic structure.}
It can seem that the absence of any special coordinate functions
on invariant phase space, that could be used to identify points of
this space, makes this space an empty abstraction.

But this is not so. Because it is possible to show that invariant
phase space, like usual one, possesses symplectic structure.

When we state one-to-one correspondence of points of the invariant
phase space and points of a usual phase space, taken in a fixed
moment of time, it turns out that their symplectic structure
correspond to each other. From this it is clear that existence of
the symplectic structure on invariant phase space is a fact which
is a {\it generalization} of Liouville and Poincare theorems.

As regards Poisson brackets, they are defined through symplectic
structure. Their mathematical definition, in fact, does not
change. But so far as this definition is applied to objects of
other nature, it turns out that such Poisson brackets are a deep
generalization of the usual. For example, they can be calculated
between values of field in {\it different} moments of time.

\paragraph{Action of Poincare group.}
If the initial Lagrangian of a field is relativistic-invariant,
then the equations of motion are also relativistic-invariant. So,
under any transformation from the Poincare group a solution of the
equations of motion is transformed into solution.

In other words, the Poincare group acts on invariant phase space.

So far as the symplectic structure on invariant phase space is
defined through Lagrangian, it turns out to be invariant under the
action of the Poincare group. So, the Poincare group acts on the
invariant phase space canonically.

\paragraph{Field representations.}
For the problem of quantization we can restrict ourself with
linear fields. These fields are characterized by the property that
addition of two solutions of equations of motions is solution.
From the point of view of invariant phase space, we can say that
in this case it possesses natural linear structure.

The Poincare group preserves this linear structure. So, the
Poincare group acts on invariant phase space as a group of {\it
symplectic} transformations.

So far as for the purpose of quantization it is necessary to
analyse this action of the Poincare group by methods of group
theory, it is useful to use the following terminology. We say that
linear relativistic fields define {\it symplectic representations}
of the Poincare group. These representations (and also their
conjugate and complexified) are also called {\it field
representations}.

We will not become more profound here in the theory of field
representations. Let us just notice that this theory is quite
analogous to the Wigner-Mackey theory of unitary representations
of Poincare group, but it plays more fundamental role for the
theory of quantized fields.

\section*{Construction of quantized field}

So, now we already have all necessary structures that are used for
{\it constructing} of quantized field. Here they are: invariant
phase space of linear classical field, symplectic structure on
this space, properly classified invariant (with respect to the
Poincare group) subspaces of field representation, and the proper
set of classical field values that are ``quantized''.

In this review I omit a description of exact construction of
quantized field. From the point of view of an algebraist all
methods used for it are well known. Similar methods are used for
construction of universal enveloping algebras for Lie algebras,
for construction of Grassmanian algebras etc.

Let us give now a list of some most important properties of the
quantization under consideration.
\begin{itemize}
\item
Quantization is performed {\it constructively}. Properties of
quantized field (for example, commutation relations) are not
postulated, but follow from the construction.
\item
The construction is explicitly relativistic-invariant.
\item
The method makes possible to see, how quantum system acquires
integrals of motion connected with symmetry group, like initial
classical system. So, it becomes possible to formulate the
mathematically rigorous analog of Noether theorem.
\item
The quantization scheme naturally covers the possibility of
quantization in a space with indefinite scalar product.
\end{itemize}

\section*{Application to electromagnetic field}

The problem of quantization of electromagnetic field was one of
the main stimuli for creation of RCQ.

It is known for a long time, that for the problems of quantum
theory it is necessary to describe electromagnetic field by means
of vector potential. It is also known that an attempt of
quantization of such a vector field leads to the necessity of
consideration of indefinite scalar product in quantum space of
states.

An indefinite scalar product, in contrast to positive-definite
case, does not define topology.

This problem was just ignored up to now (in educational
literature). By analogy with some other fields, it was believed
that the space of states of quantized electromagnetic field {\it
must} be Hilbert space, at least from topological point of view.

The RCQ method has shown that this {\it is not so}.

\section*{Other applications}

The RCQ method, of course, can be applied to other fields (for
example, to scalar, to electron-positron etc.)

It is known that, for example, the scalar field has only one
quantization in Hilbert space. Such a quantization is already
constructed, of course. And the RCQ method (if we restrict ourself
with positive-definite scalar product) {\it  can not} lead to
other quantization.

Of course, the RCQ method leads in this case to equivalent
quantization. But the undoubted merit of the method is that the
construction is performed in the frame of the general scheme,
without any ``guesses'', ``classical analogies'' etc.

It happens so because the RCQ method is a rigorous {\it
mathematical construction}.



\end{document}



\title{Что такое \rql релятивистское каноническое квантование\rqr?}
\author{Д.~А.~Арбатский\footnote{ http://daarb.narod.ru/ , http://wave.front.ru/ }}
\date{\today}
\maketitle

\begin{abstract}
Цель данного обзора~--- дать максимально доступное описание схемы
квантования релятивистских полей, которая получила название {\it
релятивистское каноническое квантование} (РКК). Здесь не даётся
полного точного изложения данной схемы. Но с помощью данного
обзора любой физик, даже не являющийся специалистом по
релятивистской квантовой теории, сможет получить общее
представление о содержании РКК, о его связи с другими
существующими подходами, о его новизне и о его плодотворности.
\end{abstract}

\section*{Инвариантный гамильтонов формализм}

Общеизвестно, что для построения квантованных полей оказывается
полезным вычислять скобки Пуассона от полевых величин. А скобка
Пуассона является понятием гамильтонова формализма.

Если перед нами стоит задача создать полностью
релятивистски-инвариантную схему квантования полей, то хотелось бы
прежде всего сформулировать гамильтонов формализм в
релятивистски-инвариантной форме.

Эту задачу наука решала удивительно долго. Фактически, считалось
общепринятым, что гамильтонов формализм нельзя сформулировать в
явно релятивистски-инвариантной форме.

Однако, с развитием методов симплектической геометрии, гамильтонов
формализм удалось сформулировать на основе таких базовых понятий,
у которых обнаружились релятивистски-инвариантные аналоги. Вот эти
понятия: фазовое пространство, симплектическая структура на
фазовом пространстве, каноническое действие однопараметрической
группы временных сдвигов на фазовом пространстве.

У релятивистских полей аналогами этих понятий выступают,
соответственно: инвариантное фазовое пространство, симплектическая
структура на инвариантном фазовом пространстве, каноническое
действие группы Пуанкаре на инвариантном фазовом пространстве.

\paragraph{Инвариантное фазовое пространство.}
Точка обычного фазового пространства характеризует динамическое
состояние системы в данный момент времени. Если для каждого
начального состояния системы уравнения движения разрешимы, причём
единственным образом, то можно говорить о взаимно-однозначном
соответствии между фазовым пространством системы (для заданного
момента времени) и множеством решений уравнений движения. Вот это
множество решений уравнений движения и называется {\it
инвариантным фазовым пространством}.

Инвариантное фазовое пространство обладает естественной структурой
многообразия. Если в качестве динамической системы выступает
релятивистское поле, то в качестве возможных координатных функций
на инвариантном фазовом пространстве могут выступать значения
величины поля в фиксированных точках пространства-времени.

Таким образом, с точки зрения структуры инвариантного фазового
пространства, возможные значения поля в фиксированной точке
пространства-времени~--- лишь одна из огромного множества функций
на инвариантном фазовом пространстве.

\paragraph{Симплектическая структура.}
Отсутствие на инвариантом фазовом пространстве каких-либо
выделенных координатных функций, с которыми следовало бы связывать
точки этого пространства, казалось бы, превращает его в
бессодержательную абстракцию.

Это, однако, не так. Ибо можно показать, что инвариантное фазовое
пространство, как и обычное, обладает симплектической структурой.

При сопоставлении точек инвариантного фазового пространства с
точками обычного фазового пространства, взятого в фиксированный
момент времени, оказывается, что их симплектические структуры
переходят друг в друга. Отсюда ясно, что существование
симплектической структуры на инвариантном фазовом пространстве~---
факт, являющийся {\it обобщением} теорем Лиувилля и Пуанкаре.

Что касается скобок Пуассона~--- то они определяются через
симплектическую структуру. Математическое определение их, по сути,
не меняется. Но поскольку это определение применяется к объектам
иной природы, оказывается, что такие скобки Пуассона являются
глубоким обобщением обычных. Например, их можно вычислять между
значениями поля в {\it разные} моменты времени.

\paragraph{Действие группы Пуанкаре.}
Если исходный лагранжиан поля релятивистски-инвариантен, то и
уравнения движения релятивистски-инвариантны. То есть при любом
преобразовании из группы Пуанкаре решение уравнений движения
переходит в решение.

Иначе говоря, группа Пуанкаре действует на инвариантном фазовом
пространстве.

Поскольку симплектическая структура на инвариантном фазовом
пространстве определяется через лагранжиан, она оказывается
инвариантной относительно действия группы Пуанкаре. То есть группа
Пуанкаре действует на инвариантном фазовом пространстве
канонически.

\paragraph{Полевые представления.}
Для целей квантования можно ограничиться линейными полями. Эти
поля характеризуются тем, что при сложении двух решений уравнений
движения получается решение. С точки зрения инвариантного фазового
пространства, можно сказать, что в этом случае оно обладает
естественной линейной структурой.

Группа Пуанкаре эту линейную структуру сохраняет. Таким образом,
группа Пуанкаре действует на инвариантном фазовом пространстве как
группа {\it симплектических} преобразований.

Поскольку для целей квантования оказывается важным изучить это
действие группы Пуанкаре методами теории групп, полезной
оказывается ещё такая терминология. Говорят, что линейные
релятивистские поля задают {\it симплектические представления}
группы Пуанкаре. Эти представления (а также их сопряжённые и
комплексифицированные) называют {\it полевыми представлениями}.

Мы не будем здесь углубляться в теорию полевых представлений.
Заметим лишь, что она во многом аналогична теории Вигнера-Макки
унитарных представлений группы Пуанкаре, но при этом она играет
для теории квантованных полей более фундаментальную роль.

\section*{Конструкция квантованного поля}

Итак, теперь мы уже имеем все необходимые структуры, из которых
{\it конструируется} квантованное поле. Вот эти структуры:
инвариантное фазовое пространство линейного классического поля,
симплектическая структура на этом пространстве, подходящим образом
классифицированные инвариантные (относительно группы Пуанкаре)
подпространства полевого представления, а также подходящее
множество классических полевых величин, которые \rql
квантуются\rqr.

В данном обзоре мы опустим описание самой конструкции
квантованного поля. С точки зрения математика-алгебраиста все
используемые при этом методы хорошо известны. Подобные методы
применяются при конструировании универсальных обёртывающих алгебр
для алгебр Ли, при конструировании грассмановых алгебр и~т.~п.

Приведём здесь перечень некоторых наиболее важных свойств
обсуждаемого метода квантования.
\begin{itemize}
\item
Квантование производится {\it конструктивно}. Свойства
квантованного поля (например, коммутационные соотношения) при этом
не постулируются, а вытекают из самой конструкции.
\item
Конструкция явно релятивистски-инвариантна.
\item
Метод даёт возможность увидеть, каким образом у квантовой системы
появляются те же интегралы движения, связанные с наличием группы
симметрии, что и у исходной классической системы. Таким образом,
удаётся сформулировать математически строгий квантовый аналог
теоремы Нётер.
\item
Данная схема квантования включает в себя совершенно естественно
возможность квантования в пространстве с индефинитным скалярным
произведением.
\end{itemize}

\section*{Применение к электромагнитному полю}

Проблема квантования электромагнитного поля была одним из главных
стимулов для создания РКК.

Давно известно, что для целей квантовой теории электромагнитное
поле нужно описывать с помощью векторного потенциала. Известно
также, что при попытке квантования такого векторного поля
возникает необходимость рассматривать индефинитное скалярное
произведение в квантовом пространстве состояний.

Индефинитное скалярное произведение, в отличие от
положительно-определённого случая, не задаёт топологии.

Эта проблема до сих пор просто замалчивалась (в учебной
литературе). По аналогии с некоторыми другими полями считалось,
что пространство состояний квантованного электромагнитного поля
{\it должно} быть гильбертовым, по крайней мере с топологической
точки зрения.

Метод РКК показал, что это {\it не так}.

\section*{Другие применения}

Метод РКК, разумеется, применим и к другим полям (например, к
скалярному, к электронно-позитронному и~т.~п.).

Известно, что, скажем, скалярное поле имеет единственное
квантование в гильбертовом пространстве. Такое квантование,
конечно, уже давно построено. И метод РКК (если ограничиться
положительно-определённым скалярным произведением) {\it не может}
приводить к другому квантованию.

Конечно, метод РКК приводит в этом случае к эквивалентному
квантованию. Но несомненное достоинство метода состоит в том, что
само построение при этом происходит в рамках общей схемы, без
каких бы то ни было \rql догадок\rqr, \rql классических
аналогий\rqr{} и~т.~п.

Так происходит потому, что метод РКК является строгой {\it
математической конструкцией}.

